\documentclass[publish]{smj}

\usepackage[greek,UKenglish]{babel}

\usepackage[autostyle=false, style=english]{csquotes}
\MakeOuterQuote{"}
\usepackage{ragged2e} 
\usepackage{rotating} 
\usepackage{tabularx}
\usepackage{booktabs}
\usepackage{threeparttable}
\usepackage{multirow}
\usepackage{xcolor}
\usepackage{adjustbox}
\usepackage{algpseudocode}
\usepackage{algorithm}
\usepackage{siunitx}

\DeclareUnicodeCharacter{025B}{$\varepsilon$} 

\newcommand{\argmin}{\arg\,\min} 

\DeclareMathOperator{\softmax}{softmax}
\DeclareMathOperator{\KL}{KL}

\Author{Schyan Zafar\Affil{1} \and Geoff K. Nicholls\Affil{1}}
\AuthorRunning{Schyan Zafar \and Geoff K. Nicholls}
  
\Affiliations{

\item Department of Statistics,       
      University of Oxford,
      Oxford,
      UK

}   

\CorrAddress{Schyan Zafar, 
             Department of Statistics, \\
             24--29 St Giles',
             Oxford, 
             OX1 3LB, 
             United Kingdom}
\CorrEmail{sczafar@stats.ox.ac.uk}
\CorrPhone{(+44)\;1865\;272860}

\Title{Exploring Learning Rate Selection in Generalised Bayesian Inference using Posterior Predictive Checks}
\TitleRunning{Exploring Learning Rate Selection in GBI using PPC}

\Abstract{
Generalised Bayesian Inference (GBI) attempts to address model misspecification in a standard Bayesian setup by tempering the likelihood. The likelihood is raised to a fractional power, called the learning rate, which reduces its importance in the posterior and has been established as a method to address certain kinds of model misspecification. Posterior Predictive Checks (PPC) attempt to detect model misspecification by locating a diagnostic, computed on the observed data, within the posterior predictive distribution of the diagnostic. This can be used to construct a hypothesis test where a small $p$-value indicates potential misfit. The recent Embedded Diachronic Sense Change (EDiSC) model suffers from misspecification and benefits from likelihood tempering. Using EDiSC as a case study, this exploratory work examines whether PPC could be used in a novel way to set the learning rate in a GBI setup. Specifically, the learning rate selected is the lowest value for which a hypothesis test using the log likelihood diagnostic is not rejected at the 10\% level.
The experimental results are promising, though not definitive, and indicate the need for further research along the lines suggested here.
}

\Keywords{
likelihood tempering; model misspecification; power likelihood; power posterior
}

\begin{document}

\maketitle


\section{Introduction}

Bayesian inference has long been a cornerstone of statistical analysis. It provides a coherent framework for updating prior beliefs $\pi(\cdot)$ about model parameter(s) $\vartheta$ in light of observed data $y$. However, traditional Bayesian methods assume that the observation model $y \sim p(\cdot | \vartheta)$ is correctly specified, an assumption that is often violated in practice. Generalised Bayesian Inference (GBI) extends traditional Bayesian methods to address model misspecification, offering a more robust approach to inference when the likelihood function does not perfectly represent the true data generating process (DGP). In its simplest form, GBI works by raising the likelihood in the standard Bayes' posterior to a fractional power $\lambda \in [0,1]$ called the learning rate --- a trick known as \textit{tempering} in some contexts --- giving the generalised Bayes' posterior
\begin{equation} \label{eq:generalised_bayes_intro}
    \pi_\lambda(\vartheta | y) \propto \pi(\vartheta) p(y|\vartheta)^\lambda \text{.}
\end{equation}

It has been shown \citep{https://doi.org/10.1111/rssb.12158} that GBI with a tempered likelihood is a valid statistical procedure well supported by decision theory, and that \eqref{eq:generalised_bayes_intro} with some $\lambda < 1$ leads to more robust posterior belief updates compared to standard Bayes when the model is misspecified \citep{10.1214/17-BA1085}. The key question then is how to select an appropriate learning rate. This is a subject of ongoing research, and several methods for selecting $\lambda$ have been proposed in the literature. Some recent papers on this subject include \citet{10.1214/17-BA1085}, \citet{10.1093/biomet/asx010}, \citet{10.1093/biomet/asz006} and \citet{10.1093/biomet/asy054}, and are reviewed in \citet{10.1214/21-BA1302}. However, these methods can only be implemented efficiently for models where the posterior summaries, integrals or other quantities required by the methods can be computed quickly. For high-dimensional models where the posteriors necessitate computationally expensive Markov Chain Monte Carlo (MCMC) sampling, the methods can be inefficient or even prohibitive. There is therefore a need for new methods for selecting $\lambda$ that can be implemented efficiently in such cases.

Posterior Predictive Checks or Checking (PPC) is a classic tool used in Bayesian analysis to assess whether the posited model does a good job of capturing aspects of the distribution of the data that are important to the modeller. PPC uses the posterior distribution of the parameters to simulate replicated datasets, and hence predict the distribution of a suitable diagnostic function of the data (or a joint function of the data and parameters). This is compared to the same diagnostic function computed on the observed data to validate the model's assumptions and fit. For a well-specified model, we should not be able to tell the observed diagnostic apart from its posterior predictions. Conversely, for a misspecified model, the observed diagnostic should be an outlier within the posterior predictive distribution. This property may be used to construct a hypothesis test (notwithstanding any miscalibration) where a 1-sided $p$-value too close to 0 or 1 would indicate potential misfit. 

A diagnostic commonly used in PPC is the log likelihood, which can be interpreted as a measure of agreement between the model parameters and data. Starting from a place of known model misspecification, specifically a model that follows the data too closely (i.e. overfits), the observed log likelihood should be large compared to the posterior predictive distribution; that is, it should fall in the right tail of the distribution, as the agreement between the parameters and data should be atypically good due to overfitting. If we then temper the likelihood with a learning rate $\lambda < 1$, the observed log likelihood should shift towards the left as we reduce its importance in the posterior, ultimately becoming an outlier in the left tail as $\lambda \to 0$. Intuitively, between these extremes, there should be a region where the observed diagnostic is typical of a well-specified model. The question we explore in this paper is whether we can use this intuition to select an appropriate $\lambda$; that is, pick a learning rate for which a hypothesis test using the log likelihood diagnostic is not rejected at a suitable significance level. We find that a 10\% threshold works well in all our real-data examples. Our approach resembles \citet{Chakraborty2023}. However, to the best of our knowledge, PPC has not previously been used to guide the choice of learning rate in quite this way.

This is exploratory work in its initial stage. We do not give any theoretical guarantees, nor make any claims that the approach studied in this paper could be generalised. In fact, it is easy to construct examples where the approach does \textit{not} work, so it is certainly not universally applicable. We elaborate further on this in the discussion at the end, where we give some intuition for when the approach is valid.

This work has been motivated by the recent Embedded Diachronic Sense Change (EDiSC) model \citep{EDiSC_ZAFAR2024108011}, which is exceptionally hard to fit to the data in question, and poses a unique set of challenges. It was found experimentally that EDiSC benefitted from likelihood tempering, making it an interesting case study for GBI. However, the existing state-of-the-art approaches for tuning $\lambda$ discussed in \citet{10.1214/21-BA1302} proved particularly challenging to implement, which motivated the search for a novel and readily implementable solution. Our approach for tuning $\lambda$ is intuitive and gives promising results in this case study, though we acknowledge its limited scope. Crucially, the learning rate selected using our method gives an improvement in the model's predictive performance, in all the examples studied, compared to standard Bayes. The next stage of this work would be to investigate what other models and settings, if any, could this approach be generalised to. 

The rest of this paper is organised as follows. In Section~\ref{sec:Background} we give some background on model misspecification, GBI and PPC. In Section~\ref{sec:setting_lambda} we introduce our novel approach for selecting the learning rate using PPC. In Section~\ref{sec:model_and_inference} we describe the EDiSC model and inferential problem that motivated this work. In Section~\ref{sec:data_and_eval_framework} we describe the data used to develop and test our method, and our performance measurement criteria. In Section~\ref{sec:experiments} we present our experimental results on development and test data. Finally, in Section~\ref{sec:discussion} we conclude with a discussion of possible conditions required for our method to work. Some further insights are given in the \nameref{sec:appendix}.

\section{Background} \label{sec:Background}

In this section, we set out the notation and describe the concepts that will be required to develop our proposed method for setting the learning rate $\lambda$ using PPC.

\subsection{Model misspecification} \label{sec:Background_misspecification}

Suppose we have a parametric model $y \sim p(\cdot | \vartheta)$ for data $y = (y_1,\dots,y_D)$ depending on some (possibly multi-dimensional) parameter $\vartheta \in \Theta$ via likelihood $p(y|\vartheta)$. The data may come in pairs $(x_d,y_d)$, with covariates $x_d = (x_{d,1},\dots,x_{d,p})$ for each $d=1,\dots,D$. Conditioning on the covariates via likelihood $p(y_d|x_d,\vartheta)$ is implied in that case, but we do not explicitly show it in this paper since $x = (x_1,\dots,x_D)$ is not a random variable. In standard Bayesian inference, we place a prior $\pi(\vartheta)$ on the parameter and update our posterior beliefs about $\vartheta$ via Bayes' theorem:
\begin{equation} \label{eq:Bayes_theorem}
    \pi(\vartheta | y) \propto \pi(\vartheta) p(y|\vartheta) \text{.}
\end{equation}
In a correctly specified model, there exists a $\vartheta^* \in \Theta$ such that $y \sim p(\cdot | \vartheta^*)$ is the true DGP. Furthermore, under regularity conditions, the Bayes' posterior \eqref{eq:Bayes_theorem} converges to a normal distribution concentrated on the true $\vartheta^*$ in the limit of infinite data $D \to \infty$ by the Bernstein–von Mises theorem. This $\vartheta^*$ also coincides with the asymptotic value of the frequentist maximum likelihood estimate.

In practice, however, a statistical model is rarely well specified, since a model is by definition a simplified representation of some real-world phenomenon. In a misspecified model, say where the true DGP is $y \sim h(\cdot)$ for some $h(\cdot) \neq p(\cdot | \vartheta)$, a `true' $\vartheta^*$ does not exist. Instead, as $D \to \infty$, under regularity conditions, both the Bayes' posterior \eqref{eq:Bayes_theorem} and the maximum likelihood estimator concentrate on the \textit{pseudo-true} parameter
\begin{equation} \label{eq:theta_KL}
    \vartheta^\dagger = \argmin_{\vartheta \in \Theta} \KL \left( h(\cdot) ~\big\Vert~ p(\cdot | \vartheta) \right) \text{,}
\end{equation}
\citep{10.1214/12-EJS675} where $\KL \left( h(\cdot) ~\big\Vert~ p(\cdot | \vartheta) \right) = \int h(y) \log \left( h(y) / p(y|\vartheta) \right) \mathrm{d}y$ is the Kullback-Leibler divergence. When $\vartheta$ has some physical meaning, this $\vartheta^\dagger$ is still interpretable and useful, and tells us something about the phenomenon being studied. However, since KL divergence places a lot of importance on the tail behaviour of the distributions in question, $\vartheta^\dagger$ learnt according to \eqref{eq:theta_KL} can be sensitive to tail misspecifications, particularly if $h(\cdot)$ has heavier tails than $p(\cdot | \vartheta)$ \citep{e20060442}. In other words, the Bayes' posterior \eqref{eq:Bayes_theorem} lacks robustness and may overfit the model.
Model misspecification can, of course, be more general, including where $\vartheta^\dagger$ has no meaningful interpretation.

\subsection{Generalised Bayesian Inference} \label{sec:Background_GBI}

GBI, in its simplest form dating back to \citet{10.1111/1467-9868.00314} and \citet{10.1214/009053606000000704}, attempts to address some kinds of model misspecification by reducing the importance of the likelihood in the standard Bayes' posterior \eqref{eq:Bayes_theorem} via tempering. The likelihood is raised to a fractional power $\lambda \in [0,1]$, called the learning rate, where $\lambda = 1$ returns standard Bayes and $\lambda = 0$ gives just the prior. For any other $\lambda$, we get the generalised Bayes' posterior
\begin{equation} \label{eq:generalised_bayes}
    \pi_\lambda(\vartheta | y) \propto \pi(\vartheta) p(y|\vartheta)^\lambda \text{.}
\end{equation}
The smaller the value of $\lambda$, the slower the rate at which posterior beliefs about $\vartheta$ are updated with an increasing number of samples $D$, and vice versa. Importantly, the interpretation of the model and parameters remains unchanged with this form of GBI: only the rate of concentration of \eqref{eq:generalised_bayes} around \eqref{eq:theta_KL} is affected, and the Bernstein-von Mises theorem (under regularity conditions) still applies \citep{JMLR:v22:20-469}. Other forms of GBI extend Bayesian inference in more general and flexible ways, including loss-function-based \citep{https://doi.org/10.1111/rssb.12158} and scoring-rule-based \citep{2021arXiv210403889P} inference. See \citet[Section~1.2]{pacchiardi2022a} for a nice summary of how the generalisations fit together. These are beyond the scope of this paper.

Intuitively \citep{10.1093/biomet/92.4.765}, a $\lambda < 1$ makes the generalised Bayes' posterior \eqref{eq:generalised_bayes} more robust to inconsistencies arising from overfitting compared to \eqref{eq:Bayes_theorem} in any finite data sample, since the information coming from the data is reduced. However, too small a value can result in not learning enough about the DGP if the signal from the data gets too weak. Hence, we need to set the learning rate $\lambda$ to a value suitable for the inferential task. Common inferential tasks include prediction and `true model' recovery (in the sense of \eqref{eq:theta_KL} for example), and it is well known that an optimal model for one task is not necessarily optimal for the other.

Four recent methods for selecting the learning rate have been reviewed in \citet{10.1214/21-BA1302}. The first of these is the SafeBayes algorithm of \citet{10.1214/17-BA1085}, which involves a grid search over potential $\lambda$ values, and selects the learning rate minimising a loss function. The loss function is computed by fitting the model iteratively, adding one data point in each iteration, and calculating the cumulative ``posterior-expected posterior-randomised (log) loss'' of predicting the next outcome in each fit. This method becomes computationally infeasible when we have hundreds or thousands of data points, and model fitting is done using MCMC, with each fit potentially taking hours to run.
  
Next, the (distinct) information-matching strategies of \citet{10.1093/biomet/asx010} and \citet{10.1093/biomet/asz006} give an explicit ``oracle'' learning rate as a formula. The computation of both oracles requires evaluating expectations with respect to the true DGP and using the pseudo-true $\vartheta^\dagger$. Since both of these are unknown, they are estimated by the empirical distribution of observed data and the maximum likelihood parameter estimates respectively. In our setting, with sparse high-dimensional data and high-dimensional parameters, these quantities are very hard to estimate with the accuracy needed to make these methods work. 

Finally, the bootstrap-motivated generalised posterior calibration (GPC) algorithm of \citet{10.1093/biomet/asy054} attempts to tune $\lambda$ iteratively over model fittings so that the generalised Bayesian posterior credible sets achieve the nominal frequentist coverage probability. This method is computationally expensive, even in simple applications where posterior credible sets are readily available, since it requires multiple model fittings using bootstrapped data for each candidate learning rate. Additionally, in high-dimensional applications where the posteriors are sampled using MCMC, a prohibitively large number of samples per model fitting is required to compute accurate coverage probabilities. A GPC variant is given by \citet{2023arXiv231012882W}, and similar remarks apply.

Separately, \citet{carmona20} use the widely applicable information criterion (WAIC) \citep{watanabe2010asymptotic, vehtari2017practical}, a predictive loss, to select the learning rate. This works well when the WAIC is a reliable estimator for the expected log pointwise predictive density (ELPD), which is not the case in our setting \citep[Section~5.2]{EDiSC_ZAFAR2024108011}. 

Setting the learning rate is a subject of ongoing research, and our literature review did not find any other methods differing substantially from those discussed in the papers cited above

\subsection{Posterior Predictive Checks} \label{sec:PPCs}

PPC is a classic \citep{https://doi.org/10.1111/j.2517-6161.1967.tb00676.x, 10.1214/aos/1176346785} diagnostic tool used frequently in Bayesian analysis to assess a model's ability to describe the data. Given observed data $y^\text{obs} = (y_1^\text{obs},\dots,y_D^\text{obs})$, PPC works by simulating $y^{\text{rep}} \sim p(\cdot | y^\text{obs})$ from the posterior predictive distribution
\begin{equation} \label{eq:post_pred_dist}
    p(\cdot | y^\text{obs}) = \int_{\Theta} p(\cdot | \vartheta) \pi(\vartheta | y^\text{obs}) \,\mathrm{d} \vartheta 
\end{equation}
to give replicated data $y^{\text{rep},n}, n=1,\dots,N$, where each replicate $y^{\text{rep},n} = (y_1^{\text{rep},n},\dots,y_D^{\text{rep},n})$ is a dataset of the same size as $y^\text{obs}$, using the same covariates $x$ (if applicable).
In a basic PPC, a diagnostic function $s(y)$ of the data is chosen, which may be some quantity of interest to the modeller such as the sample mean $\bar{y} = \frac{1}{D} \sum_{d=1}^D y_d$ or variance $\frac{1}{D-1} \sum_{d=1}^D \left( y_d - \bar{y} \right)^2$, to investigate the compatibility between the data $y$ and the model. The diagnostic is computed on both observed and replicated data to give, respectively, the observed statistic $s(y^\text{obs})$ and samples $s(y^{\text{rep},1}), \dots, s(y^{\text{rep},N})$ from the \textit{reference distribution} of $s(y^\text{rep})$. The idea is that, if the model is true,  has a mixture distribution over $\vartheta \in \Theta$ with a component $\vartheta=\vartheta^*$ that matches the generative model for $s(y^\text{obs})$. Conversely, if $s(y^\text{obs})$ is an outlier within the distribution of $s(y^\text{rep})$, it is indicative of a misspecified model. 

\citet{10.1214/aos/1176325622} and \citet{62bfc978-09b1-3997-9776-380d0b45e9c2} extend the basic PPC to use a ``realised discrepancy'' diagnostic function $s(y, \vartheta)$, which depends on both the data and the model parameters, and quantifies the discrepancy between them. This could, for instance, be the negative log likelihood $-\sum_{d=1}^D \log p(y_d | \vartheta)$ or the chi-squared diagnostic $\frac{1}{D} \sum_{d=1}^D \frac{(y_d - \mathbb{E}(y_d|\vartheta))^2}{\mathbb{V}(y_d|\vartheta)}$. Quoting \citet[Section~2.2]{62bfc978-09b1-3997-9776-380d0b45e9c2}, ``the focus here is to measure discrepancies between a model and the data, not to test whether a model is true''. The idea is conceptually the same as for $s(y)$: we locate $s(y^\text{obs}, \vartheta)$ within the reference distribution of $s(y^\text{rep}, \vartheta)$ and check whether it is an outlier. Note that, in this paper, we take the diagnostic to be the (positive) log likelihood, so $s(y, \vartheta)$ is interpreted as the \textit{agreement}, rather than the discrepancy, between $y$ and $\vartheta$.

A common way to check whether $s(y^\text{obs}, \vartheta)$ is an outlier within the distribution of $s(y^\text{rep}, \vartheta)$ is to use a hypothesis test with a 1-sided `$p$-value' 
\begin{equation} \label{eq:ppc_pvalue1}
    p_\text{ppc} = \mathbb{P} \left( s(y^\text{rep}, \vartheta) < s(y^\text{obs}, \vartheta) \mid y^\text{obs} \right) \text{,}
\end{equation}
but other measures of surprise \citep{BAYARRI20033} or visual approaches \citep{62bfc978-09b1-3997-9776-380d0b45e9c2, doi:10.1198/106186004X11435} may also be used. The direction of the inequality in \eqref{eq:ppc_pvalue1} may be reversed depending on the specific diagnostic $s(y,\vartheta)$ chosen in practice, or indeed a 2-sided `$p$-value' may be more appropriate. We write `$p$-value' within quotes since it is well known \citep{10.1214/aos/1176325622, 4ee526ff-bb22-3910-b521-ec4545bc217a, 075ca5bb-7cd5-3c54-8f3d-5c156c8eb895} that this hypothesis test is in general not calibrated; that is, under the null hypothesis of a well-specified model, the distribution of $p_\text{ppc}$ is not uniform over $[0,1]$. The miscalibration results from the double use of the data: we use $y^\text{obs}$ to generate replicated data in \eqref{eq:post_pred_dist}, and the same data to compute $p_\text{ppc}$ in \eqref{eq:ppc_pvalue1}. 

A number of alternatives and solutions have been proposed in the literature to address the miscalibration. These include post-hoc corrections \citep{075ca5bb-7cd5-3c54-8f3d-5c156c8eb895, doi:10.1198/016214505000001393}, partial posterior predictive and conditional predictive $p$-values \citep{4ee526ff-bb22-3910-b521-ec4545bc217a}, split predictive checks \citep{2022arXiv220315897L}, and holdout predictive checks \citep{10.1093/jrsssb/qkad105} among others. \citet[Sections 1.1~\&~4]{10.1093/jrsssb/qkad105} give a brief summary and criticism of all these methods. Nevertheless, despite being miscalibrated, it is still the case that a value of $p_\text{ppc}$ in \eqref{eq:ppc_pvalue1} too close to 0 or 1 would potentially indicate misfit. The only consequence of the miscalibration is that, if we reject the null hypothesis of a well-specified model at significance level $\alpha$, the probability of a type~I error will usually be smaller than $\alpha$ \citep{10.1214/aos/1176325622}. 

It is, of course, not possible to compute \eqref{eq:ppc_pvalue1} directly, since the parameter $\vartheta$ is unknown. However, it is easy to estimate \eqref{eq:ppc_pvalue1} using Monte Carlo methods --- a property that makes PPC a natural and computationally efficient diagnostic tool to use when the posterior is sampled with MCMC. Given posterior samples $\vartheta^n \sim \pi(\cdot | y^\text{obs}), n = 1,\dots,N,$ we first simulate a replicated dataset $y^{\text{rep},n} \sim p(\cdot | \vartheta^n)$ for each $n$ using the posited model. Then, for each pair $(y^{\text{rep},n}, \vartheta^n)$, we compute $s(y^{\text{rep},n}, \vartheta^n)$ to get the reference empirical distribution of the diagnostic. 

For the observed diagnostic, there are a few options. One option is to use the same posterior samples $\vartheta^n$ to compute $s(y^{\text{obs}}, \vartheta^n)$ for each $n$. This gives us a diagnostic pair (replicated and observed) for each posterior sample, and we can compute
\begin{equation} \label{eq:ppc_pvalue2}
    p_\text{ppc} = \frac{1}{N} \sum_{n=1}^N \mathbb{I} \left( s(y^{\text{rep},n}, \vartheta^n) < s(y^{\text{obs}}, \vartheta^n) \right) \text{.}
\end{equation}
This option is favoured by \citet{62bfc978-09b1-3997-9776-380d0b45e9c2}, who also recommend making a scatter plot of $s(y^{\text{obs}}, \vartheta^n)$ against $s(y^{\text{rep},n}, \vartheta^n)$ for visual inspection. A second option is to use an average $\bar{s}(y^{\text{obs}})$, such as the mean $\frac{1}{N} \sum_{n=1}^N s(y^{\text{obs}}, \vartheta^n)$ or ${\text{median}_{n}} s(y^{\text{obs}}, \vartheta^n)$ over $n \in \{1,\dots,N\}$, to approximate $s(y^\text{obs}, \vartheta)$, and then compute
\begin{equation} \label{eq:ppc_pvalue3}
    p_\text{ppc} = \frac{1}{N} \sum_{n=1}^N \mathbb{I} \left( s(y^{\text{rep},n}, \vartheta^n) < \bar{s}(y^{\text{obs}}) \right) \text{.}
\end{equation} 
A third option is to use a summary statistic such as the posterior mean $\bar{\vartheta} = \frac{1}{N} \sum_{n=1}^N \vartheta^n$ to approximate $\vartheta$, and use this to compute
\begin{equation} \label{eq:ppc_pvalue4}
    p_\text{ppc} = \frac{1}{N} \sum_{n=1}^N \mathbb{I} \left( s(y^{\text{rep},n}, \vartheta^n) < s(y^{\text{obs}}, \bar{\vartheta}) \right) \text{.}
\end{equation}
We will use the third option in our experiments for the reasons given below. 

If the posterior distribution is concentrated on the pseudo-true \eqref{eq:theta_KL} then the three $p$-values above will be similar. However, in practice, the regularity conditions for the Bernstein-von Mises theorem do not always hold, and in any case the convergence to \eqref{eq:theta_KL} may be extremely slow. Thus, with finite data, the posterior samples $\vartheta^1, \dots, \vartheta^N$ could be far from the pseudo-true $\vartheta^\dagger$, especially in high dimensions. This is indeed the case for all real-data examples studied in this paper. Even if the model is well specified, the diffuseness of the posterior results in the agreement $s(y^{\text{obs}}, \vartheta^n)$ being deflated by the parameter variance, and typically much lower than $s(y^{\text{obs}}, \vartheta^*)$. We wish to remove this parameter variance.

If the model is correct then the agreement between the observed data and the true parameter should be about the same as the agreement between simulated data and the parameter at which it was simulated,
so $s(y^{\text{obs}},\vartheta^*)$ and $s(y^{\text{rep},n},\vartheta^n)$ should have similar magnitude. We assume that the posterior mean $\bar{\vartheta}$ is a reasonable estimate for the true parameter $\vartheta^*$, and test a null hypothesis of a well-specified model using the $p$-value in \eqref{eq:ppc_pvalue4} rather than \eqref{eq:ppc_pvalue2} or \eqref{eq:ppc_pvalue3}.
This is quite a strong assumption in our setting. After all, we are interested in correcting for misspecification, and this may bias the posterior mean. However, as we shall see in Section~\ref{sec:experiments}, our approach for correcting misspecification (set out in Section~\ref{sec:setting_lambda} below) seems quite effective, which suggests that misspecification is causing under-dispersion rather than biasing the mean.

\section{Setting $\lambda$ using PPC} \label{sec:setting_lambda}

We now describe our proposed method for setting the learning rate $\lambda$ using PPC, which bears strong resemblance to the approach taken by \citet[Section~4.2]{Chakraborty2023} to set the influence parameter in semi-modular inference (\citealt{carmona20}, a variant of generalised Bayes). We start with observed data $y^\text{obs} = (y_1^\text{obs},\dots,y_D^\text{obs})$, parametric model $y^\text{obs} \sim p(\cdot | \vartheta)$, prior $\pi(\vartheta)$, and candidate learning rates $\lambda_r, r=1,\dots,R,$ including $\lambda_1 = 1$. We also choose a diagnostic function $s(y, \vartheta)$ and a significance level $\alpha$ appropriate for the application. For each $\lambda \in \{\lambda_1,\dots,\lambda_R\}$, we target the generalised Bayes' posterior
\begin{equation} \label{eq:generalised_bayes_lambda}
    \pi_\lambda(\vartheta | y^\text{obs}) \propto \pi(\vartheta) p(y^\text{obs}|\vartheta)^\lambda
\end{equation}
to obtain posterior samples $\vartheta^{\lambda, n} \sim \pi_\lambda(\cdot | y^\text{obs})$, $n = 1,\dots,N$. Using these, we generate replicated data $y^{\text{rep},\lambda,n} \sim p(\cdot | \vartheta^{\lambda, n})$ to perform a PPC and compute the $p$-value as in \eqref{eq:ppc_pvalue4}, denoted
\begin{equation} \label{eq:ppc_pvalue_lambda}
    p_\text{ppc}^\lambda = \frac{1}{N} \sum_{n=1}^N \mathbb{I} \left( s(y^{\text{rep},\lambda,n}, \vartheta^{\lambda, n}) < s(y^{\text{obs}}, \bar{\vartheta}^\lambda) \right) \text{,}
\end{equation}
where $\bar{\vartheta}^\lambda = \frac{1}{N} \sum_{n=1}^N \vartheta^{\lambda,n}$.  Then, we select the rate minimising \eqref{eq:ppc_pvalue_lambda} subject to $p_\text{ppc}^\lambda > \alpha$, that is
\begin{equation} \label{eq:selected_lambda}
    \lambda^\dagger = \argmin_{\lambda \in \{\lambda_1,\dots,\lambda_R\}} \{p_\text{ppc}^\lambda: p_\text{ppc}^\lambda > \alpha\} \text{.}
\end{equation}
The method is summarised in Algorithm~\ref{alg:set_lambda_using_ppc}.

\begin{algorithm}[!t]
\caption{Setting the generalised Bayes' learning rate using Posterior Predictive Checks}
\label{alg:set_lambda_using_ppc}
\begin{algorithmic}[1]
\State for observed data $y^\text{obs} = (y_1^\text{obs},\dots,y_D^\text{obs})$, specify model $y^\text{obs} \sim p(\cdot | \vartheta)$ and prior $\pi(\vartheta)$
\State specify candidate learning rates $\lambda_1,\dots,\lambda_R$, with $\lambda_1 = 1$ 
\State choose diagnostic function $s(y, \vartheta)$ and significance level $\alpha$
\For {learning rate $\lambda \in \{\lambda_1,\dots,\lambda_R\}$}
    \State run MCMC targeting $\pi_\lambda(\vartheta | y^\text{obs}) \propto \pi(\vartheta) p(y^\text{obs}|\vartheta)^\lambda$
    \State obtain posterior samples $\vartheta^{\lambda, n} \sim \pi_\lambda(\cdot | y^\text{obs})$, $n = 1,\dots,N$
    \For {$n \in \{1,\dots,N\}$}
        \State simulate replicated dataset $y^{\text{rep},\lambda,n} \sim p(\cdot | \vartheta^{\lambda, n})$ and evaluate diagnostic $s(y^{\text{rep},\lambda,n}, \vartheta^{\lambda, n})$
    \EndFor
    \State compute posterior mean $\bar{\vartheta}^\lambda = \frac{1}{N} \sum_{n=1}^N \vartheta^{\lambda,n}$ and evaluate observed diagnostic $s(y^{\text{obs}}, \bar{\vartheta}^\lambda)$
    \State compute $p$-value $p_\text{ppc}^\lambda = \frac{1}{N} \sum_{n=1}^N \mathbb{I} \left( s(y^{\text{rep},\lambda,n}, \vartheta^{\lambda, n}) < s(y^{\text{obs}}, \bar{\vartheta}^\lambda) \right)$
\EndFor
\State set learning rate $\lambda^\dagger = \argmin_{\lambda \in \{\lambda_1,\dots,\lambda_R\}} \{p_\text{ppc}^\lambda: p_\text{ppc}^\lambda > \alpha\}$
\end{algorithmic}
\end{algorithm}

The direction of the inequality in \eqref{eq:ppc_pvalue_lambda} assumes that we have chosen a diagnostic $s(y,\vartheta)$ where a large value indicates greater agreement between the data and parameters, such as the log likelihood $\log p(y|\vartheta)$. If the opposite is true, we could simply take the negative function as our diagnostic. This convention helps us to think in log-likelihood terms, which is the chosen diagnostic in our application, rather than discrepancy terms (as is usual with PPC).

Intuitively, our method works on the premise that, at $\lambda=1$, we are overfitting the model; so the agreement is high and $s(y^{\text{obs}}, \bar{\vartheta}^\lambda)$ must lie in the right tail of the reference distribution of $s(y^{\text{rep},\lambda}, \vartheta^{\lambda})$. Then, as we reduce $\lambda$, the generalised Bayes' posterior \eqref{eq:generalised_bayes_lambda} tracks the data less closely, which reduces the agreement and moves $s(y^{\text{obs}}, \bar{\vartheta}^\lambda)$ to the left relative to the reference distribution. Reducing $\lambda$ reduces the amount of information coming from the data as it shifts the weight in \eqref{eq:generalised_bayes_lambda} from $p(y^\text{obs}|\vartheta)$ to $\pi(\vartheta)$. This makes sense when the data is actually correlated but the fitted model assumes independence, since then the effective sample size of the data is a lot smaller than its nominal sample size, and removing information from the data is a good thing. 
However, this makes the fit worse. Our method implicitly assumes that this tradeoff is desirable up to the point that the agreement $s(y^{\text{obs}}, \bar{\vartheta}^\lambda)$ is representative of a well-specified model. We therefore stop at the point beyond which the PPC would reject the model as misspecified.

The significance level $\alpha$ in a hypothesis test is always subjective, but 0.05 is usually a popular choice. However, as noted in Section~\ref{sec:PPCs}, the $p_\text{ppc}^\lambda$ in \eqref{eq:ppc_pvalue_lambda} is not a calibrated $p$-value. It is, in fact, biased towards 0.5 \citep{075ca5bb-7cd5-3c54-8f3d-5c156c8eb895}, and the true probability of a Type~I error is typically less than $\alpha$ \citep{10.1214/aos/1176325622}. Therefore, as a rough guide, if we wish to target a Type~I error probability of 0.05, we should set $\alpha$ to a value higher than 0.05. In our exploratory analysis, we find that $\alpha = 0.1$ serves us well.

For the candidate values $\lambda_r, r=1,\dots,R,$ we could set these to $\lambda_r = 2^{-(r-1)\kappa}$ for some $\kappa$ as in \citet{10.1214/17-BA1085}. Alternatively, we could set these at suitably small fixed-size decrements from 1. We could also proceed heuristically, and tune $\lambda$ to zero-in on $p_\text{ppc}^\lambda = \alpha$, rather than doing a grid search. The exact choice is not that important, and depends on the application as well as computational resources available. In our exploratory analysis, we use decrements of 0.1, which is sufficient for our purpose.

\section{Model and inference} \label{sec:model_and_inference}

In this section, we describe the model and inferential problem that motivated this work. The recent EDiSC model \citep{EDiSC_ZAFAR2024108011}, like its predecessor models DiSC \citep{DiSC_https://doi.org/10.1111/rssc.12591}, GASC \citep{perrone-etal-2019-gasc} and SCAN \citep{frermann2016bayesian}, is used to infer and analyse the evolving meanings or senses of a given target word in an unsupervised setting. It is a bag-of-words model in which grammar and syntax are ignored, and is closely related to topic models \citep{Blei:2006:DTM:1143844.1143859, dieng2019dynamic, 10.1162/tacl_a_00325}. The model is fitted to a set of text snippets centred on the target word. It exploits the idea that context words inform the sense of the target word in any given snippet. An example is the word ``bank'' with two distinct senses of riverbank and financial institution. Context words like ``water'' or ``stream'' would indicate the former sense, whereas context words like ``money'' or ``finance'' would indicate the latter. Another example for the word ``bug'' is given in Table~\ref{tab:bug_snippets}. We can thus posit a generative model where context words are sampled conditional on the target-word sense. If the data spans multiple time periods, a time dimension in the model captures the diachronic sense change. 

\begin{table*}[!t]
\caption{Example text snippets for target word ``\textcolor{red}{bug}'' showing its four different senses. Context words are lemmatised, and \textcolor{blue}{stopwords}, \textcolor{orange}{infrequent words} and punctuation are dropped, to get the data used in model fitting.} \label{tab:bug_snippets}
\centering
\begin{tabularx}{\textwidth}{p{1.5cm} X}
\toprule
insect & ... insect repellent \textcolor{blue}{on a} winter trip \textcolor{blue}{when there are no} \textcolor{red}{bugs} around. \textcolor{blue}{Your} first-aid kit \textcolor{blue}{should} reflect \textcolor{blue}{your} personal needs \textcolor{blue}{as} ...\\
\midrule
micro-organism & ... \textcolor{blue}{These} intruders \textcolor{blue}{are what} cause \textcolor{blue}{the} fever, \textcolor{blue}{for the} TB \textcolor{red}{bugs} \textcolor{blue}{are not} virulent enough \textcolor{blue}{to} cause high temperatures. \textcolor{blue}{The} effect ...\\
\midrule
software glitch & ... bug \textcolor{blue}{the} Quality Assurance people find \textcolor{blue}{and} \textcolor{orange}{\$20} \textcolor{blue}{for each} \textcolor{red}{bug} \textcolor{blue}{the} programmers fix. \textcolor{blue}{These are the same} programmers \textcolor{blue}{who} create ...\\
\midrule
tapping device & ... \textcolor{blue}{too} much information \textcolor{blue}{has been} collected \textcolor{blue}{through} secret informants, wiretaps, \textcolor{red}{bugs}, surreptitious mail opening \textcolor{blue}{and} break-ins, \textcolor{blue}{the} Church Report \textcolor{blue}{had} warned ...\\
\bottomrule
\end{tabularx}
\end{table*}

For a detailed exposition of the model and notation, we refer the reader to \citet[Sections 2~\&~4]{EDiSC_ZAFAR2024108011}. Here, we summarise only the elements required for the current paper. The data $W$ for a target word consists of $D$ snippets, and spans $T$ discrete contiguous time periods and $G$ genres. Snippets are a fixed window of $L$ lemmatised context words around the target word; but stopwords and very low frequency words are filtered out, leaving some context positions empty. Each snippet $W_d, d = 1,\dots,D,$ belongs to time period $\tau_d$ and genre $\gamma_d$. In the generative model, for each $W_d$, we first sample its sense $z_d$, and then sample context words given the sense. Sense $z_d$ is sampled from a multinomial sense-prevalence distribution $\tilde{\phi}^{\gamma_d,\tau_d}$ over $K$ senses (indexed by genre and time) so that $p(z_d=k | \tilde{\phi}^{\gamma_d,\tau_d}) = \tilde{\phi}^{\gamma_d,\tau_d}_k$ for all senses $k \in \{1,\dots,K\}$. Given $z_d$, at each position $i$ in $W_d$, context words $w_{d,i}$ are sampled independently from a multinomial distribution $\tilde{\psi}^{z_d,\tau_d}$ over the $V$-sized lemmatised vocabulary (indexed by sense and time) so that $p(w_{d,i}=v | z_d, \tilde{\psi}^{z_d,\tau_d}) = \tilde{\psi}^{z_d,\tau_d}_v$ for all words $v \in \{1,\dots,V\}$. The inferential task is to learn $\tilde{\phi}$ and $\tilde{\psi}$ given $W$.

We use the notation $(z_d, \gamma_d, \tau_d)$ for snippet-specific sense-genre-time triples, and $(k,g,t)$ for their generic equivalents. We define $\tilde{\phi}^{g,t} = \softmax ({\phi}^{g,t})$ and $\tilde{\psi}^{k,t} = \softmax ({\psi}^{k,t})$, and place priors on the real arrays $\phi$ and $\psi$ as described in \citet[Sections~4]{EDiSC_ZAFAR2024108011}. In the notation of Section~\ref{sec:Background_misspecification}, we thus have $\vartheta = (\phi, \psi)$ and $y=(W,z)$. Each data `point' $y_d = (W_d, z_d)$ is a composite made up of an independent sample $z_d$ and a set of conditionally independent samples $w \in W_d$, and has deterministic covariates $x_d = (\gamma_d, \tau_d)$. Snippets $W$ are observed data, whereas sense assignments $z = (z_1,\dots,z_D)$ are missing data. Taking a Bayesian approach, $z$ is treated as a latent variable and marginalised out analytically to give the likelihood
\begin{equation} \label{eq:likelihood}
    p(W | \phi, \psi) = \prod_{d=1}^D \sum_{k=1}^K p(z_d=k | \phi) p(W_d | z_d=k, \psi) = \prod_{d=1}^D \sum_{k=1}^K \tilde{\phi}_k^{\gamma_d,\tau_d} \prod_{w \in W_d} \tilde{\psi}_{w}^{k,\tau_d} \text{.}
\end{equation}

It is easy to see why this model is inherently misspecified. In any natural language, syntax and grammar are important, and context words are neither independent nor random. Target-word sense depends on more than just the time and genre in reality. Snippets are not actually generated in isolation, but are found within a larger document or corpus. Lemmatising the vocabulary and filtering out stopwords and very low frequency words is a gross simplification of actual usage. Furthermore, each word in the vocabulary is represented by an $M$-dimensional embedding vector (learnt independently of the snippet data), and there would inevitably be further misspecifications in the embedding model. 

Notwithstanding the above, in most real cases, there is at least some dependence between any target-word sense and the context words used for that sense. Also, `physical' parameters retain their meaning under misspecification: there must always be some actual sense-prevalence frequency. It follows that the inferred $\tilde{\psi}$ and $\tilde{\phi}$ are well defined despite the misspecification. Our goal, then, is to learn the parameters for which the posited model \eqref{eq:likelihood} is closest to the latent `true' DGP in the sense of \eqref{eq:theta_KL}. 

There are, in the fitted model, many assumptions of independence, whereas in fact there must be very complex correlations. Fitting a model that assumes independence to correlated data often gives credible sets that are too small. However, a realistic parametric model for these complex correlations seems infeasible, and non-parametric approaches are unlikely to help due to the sparsity of data. Faced with this, we turn to GBI: instead of elaborating the model, we modify the inference framework and target the Generalised Bayes' posterior
\begin{equation} \label{eq:generalised_bayes_EDiSC}
    \pi_\lambda(\phi, \psi | W) \propto \pi(\phi, \psi) p(W|\phi,\psi)^\lambda \text{.}
\end{equation}
The posterior \eqref{eq:generalised_bayes_EDiSC} suffers from ridge structures and multimodality, making simple variational methods unreliable \citep[Section~4.4]{EDiSC_ZAFAR2024108011}. The dimension is too high for quadrature, so we are restricted to MCMC methods. For efficiency reasons, we use the No-U-Turn sampler (NUTS, \citealt{hoffman2014no}) from the Stan software \citep{Stan_citation, RStan_citation} for all MCMC used in this paper.

\section{Data and evaluation framework} \label{sec:data_and_eval_framework}

Here we describe the data used to develop and test our method, as well as the criteria used for measuring performance.

\subsection{Development and test data} \label{sec:dev_and_test_data}

\citet[Section~2]{EDiSC_ZAFAR2024108011} describe the four datasets used in their experiments. Three of these --- the main focus of their paper --- come from the Diorisis Ancient Greek Corpus \citep{TheDiorisisAncientGreekCorpus}, with expert sense-annotation provided by \citet{vatri_lahteenoja_mcgillivray_2019}. These are for the target words ``kosmos'' (with decoration, order and world senses), ``mus'' (with mouse, muscle and mussel senses) and ``harmonia'' (with abstract, concrete and musical senses). We use these three datasets for exploratory analysis and to develop our method for setting the learning rate $\lambda$ given in Section~\ref{sec:setting_lambda}. Since we explored a number of $\lambda$-selection methods (which we do not report) on these data, it would be misleading to evaluate the performance of our chosen method on just these data. We therefore test our method on the additional datasets described below.

The fourth dataset used by \citet{EDiSC_ZAFAR2024108011} is for the target word ``bank'' from the Corpus of Historical American English (COHA, \citealt{davies2010corpus, 9789027273192-cilt.325.11dav}). The authors manually annotated 3,525 snippets for ``bank'' with the sense of riverbank or financial institution. We split this data into five roughly equal-sized subsets to increase the datasets available for testing. For the ``bank'' and ancient Greek data, we use the same modelling choices (number of senses $K$ and embedding dimension $M$) as \citet[Section~5.2]{EDiSC_ZAFAR2024108011}.

In addition to these datasets, we extract five new datasets from the Clean Corpus of Historical American English (CCOHA, \citealt{alatrash-etal-2020-ccoha}), an updated version of COHA with fewer typographical errors, and manually annotate the snippets with the correct sense. These are for the target words ``chair'' (chairman or furniture), ``apple'' (Apple Inc. company or fruit), ``gay'' (bright/showy, forward/bold, merry/lively/social or homosexual), ``mouse'' (computer pointing device or rodent) and ``bug'' (insect, microorganism, software glitch or tapping device). The ``chair'', ``gay'' and ``bug'' data are obtained from the corpus using stratified random sampling over genres and time periods, whereas the ``apple'' and ``mouse'' samples are selected to ensure both senses are represented adequately. Embeddings of dimension $M=200$ are learnt for all context words in the corpus with minimum frequency 10 using GloVe \citep{pennington2014glove}. Compared to the earlier ``bank'' data, we retain a larger vocabulary in the filtering process so as not to lose important semantic information, but consequently also retain more noise.

\begin{table}[t]
\caption{Data summary} \label{tab:data_summary}
\centering
\begin{adjustbox}{max width=\textwidth}
\begin{tabular}{c lrrcccclrl}
\toprule
\multirow{2}{*}{\rotatebox{90}{Corpus}} &        & Snippets & Vocab & Length & True senses & Model senses & \multicolumn{2}{l}{Genres} & \multicolumn{2}{l}{Time periods} \\
&Target word & \multicolumn{1}{c}{($D$)} & \multicolumn{1}{c}{($V$)} & ($L$) & ($K'$) & ($K$) & ($G$) & detail & ($T$) & detail \\ \\[-1.25em]
\midrule
\multirow{3}{*}{\rotatebox{90}{Diorisis}} &kosmos   & 1,469\phantom{N} & 2,904 & 14 & 3 & 4 & 2 & \multirow{1}{*}{\shortstack[l]{narrative, non-narr}} & 9 & 700~BC to 200~AD, centuries \\
&mus      &   214\phantom{N} &   899 & 14 & 3 & 3 & 2 & \multirow{2}{*}{\shortstack[l]{technical and\\ non-technical}} & 9 & 500~BC to 400~AD, centuries \\
&harmonia &   653\phantom{N} & 1,607 & 14 & 3 & 4 & 2 & & 12 & 800~BC to 400~AD, centuries \\ \cline{1-1}
\multirow{5}{*}{\rotatebox{90}{COHA}} &bank split 1 &   704\phantom{N} &   736 & 14 & 2 & 2 & 1 & \multirow{5}{*}{\shortstack[l]{merged news,\\ magazine\\ fiction, non-fic}} & 10 & 1810--2010, 20yr intervals \\
&bank split 2 &   708\phantom{N} &   717 & 14 & 2 & 2 & 1 & & 10 & 1810--2010, 20yr intervals \\
&bank split 3 &   703\phantom{N} &   728 & 14 & 2 & 2 & 1 & & 10 & 1810--2010, 20yr intervals \\
&bank split 4 &   704\phantom{N} &   742 & 14 & 2 & 2 & 1 & & 10 & 1810--2010, 20yr intervals \\
&bank split 5 &   706\phantom{N} &   735 & 14 & 2 & 2 & 1 & & 10 & 1810--2010, 20yr intervals \\ \cline{1-1}
\multirow{5}{*}{\rotatebox{90}{CCOHA}} &chair   &   745\phantom{N} & 3,180 & 20 & 2 & 2 & 4 & \multirow{2}{*}{\shortstack[l]{news, magazine,\\ fic, non-fic/acad}} & 10 & 1820--2020, 20yr intervals \\
&apple   & 1,154\phantom{N} & 3,737 & 20 & 2 & 2 & 4 & & 6 & 1960--2020, 10yr intervals \\
&gay   &     650\phantom{N} & 3,071 & 20 & 2 & 4 & 1 & \multirow{3}{*}{\shortstack[l]{merged news,\\ magazine,\\ non-fic/acad}} & 5 & 1920--2020, 20yr intervals \\
&mouse &     584\phantom{N} & 2,439 & 20 & 2 & 3 & 1 & & 4 & 1940--2020, 20yr intervals \\
&bug   &     522\phantom{N} & 2,475 & 20 & 4 & 4 & 1 & & 8 & 1980--2020, 5yr intervals \\
\bottomrule
\end{tabular}
\end{adjustbox}
\end{table}

The datasets are summarised in Table~\ref{tab:data_summary}. Note that the manual sense annotations are used only in the evaluation and not in model fitting. The notation $K'$ refers to the number of true target-word senses, whereas the models are fitted using $K$ senses, which may be different to $K'$. Model selection with respect to the choice of $K$ is done as per \citet[Section~5.2]{EDiSC_ZAFAR2024108011}. For target word ``gay'', during the manual annotation, it was often quite difficult for us to identify the correct sense out of the three non-homosexual senses owing to the very subtle differences between them; hence, for evaluation purposes, we combined the three senses into a single non-homosexual sense.

\subsection{Assessing model performance} \label{sec:assessing_performance}

\citet[Section~5]{EDiSC_ZAFAR2024108011} assess model performance on two fronts: predictive accuracy and true-model recovery. The true model is obviously unknown, but a proxy `true' model is used instead. Given true sense assignments $o = (o_1,\dots,o_D)$, the standard Bayes' posterior conditioned on the truth, $\tilde{\phi}|(z=o)$, is used as a well-calibrated independent estimate for the unknown true sense prevalence $\tilde{\Phi}$; and the model posterior $\tilde{\phi}|W$ is compared against it. Then, the overlap between $\tilde{\phi}|(z=o)$ and $\tilde{\phi}|W$ quantifies performance on true-model recovery. However, using this comparison in a GBI setup would not be appropriate since, arguably, we could have a generalised Bayes' posterior $\tilde{\phi}^\lambda|(z=o)$, for some $\lambda < 1$, that is in fact closer to $\tilde{\Phi}$ than $\tilde{\phi}|(z=o)$; and finding a suitable $\lambda$ value is the very problem we are addressing. On the other hand, assessing predictive accuracy only requires the objective truth $o = (o_1,\dots,o_D)$, and is therefore a more suitable measure of performance.

Following \citet[Section~5.1]{EDiSC_ZAFAR2024108011}, we assess predictive accuracy under each learning rate $\lambda$ using the Brier score
\begin{equation} \label{eq:Brier_score_lambda}
    \text{BS}_\lambda = \frac{1}{D} \sum_{d=1}^{D} \sum_{k=1'}^{K'} \left( \hat{p}_\lambda (z_d=k) - \mathbb{I} (o_d=k) \right)^2 \text{,}
\end{equation}
a proper scoring rule for multi-category probabilistic predictions $\hat{p}_\lambda (z_d=k)$, ranging from 0 (best) to 2 (worst). Here, $\hat{p}_\lambda (z_d=k)$ is the estimated value of $\mathbb{E}_{(\phi,\psi)^\lambda|W} \big( p(z_d=k | W_d, \phi, \psi) \big)$, computed on the MCMC output $(\phi,\psi)^\lambda|W$ targeting the generalised Bayes' posterior \eqref{eq:generalised_bayes_EDiSC} with learning rate $\lambda$. The estimate is obtained by normalising 
\begin{equation} \label{eq:z_d_posterior}
    p(z_d=k | W_d, \phi, \psi) \propto \tilde{\phi}^{\gamma_d,\tau_d}_{k} \prod_{w \in W_d} \tilde{\psi}^{k,\tau_d}_{w}
\end{equation}
over senses $k \in \{1,\dots,K\}$ for each posterior sample, and then averaging the normalised \eqref{eq:z_d_posterior} across samples. Recall that we have $K'$ true senses, whereas we run the models using $K$ senses (with $K \geq K'$), so modelled senses may be grouped together to map them onto the true senses. 

Within the context of this exploratory work, we would ideally like to select the `optimal' learning rate $\lambda^* = \argmin_\lambda \text{BS}_\lambda$. However, unless we use the true sense assignments in some way, this can only be coincidental, since GBI attempts only to correct model misspecification in the sense implicit in Section~\ref{sec:setting_lambda} rather than specifically find the best predictive model. We expect it to do reasonably well, as our criterion for selecting $\lambda$ should improve posterior calibration, and the Brier score is sensitive to calibration (among other factors). Therefore, as long as the $\lambda^\dagger$ selected as per \eqref{eq:selected_lambda} gives a Brier score $\text{BS}_{\lambda^\dagger} \le \text{BS}_1$ and reasonably close to the optimal $\text{BS}_{\lambda^*}$, that would indicate success.

\section{Experiments} \label{sec:experiments}

We describe our experiments on development data, followed by the results on test data.

\subsection{Method development} \label{sec:method_development}

Here we give the results of experiments performed on the three ancient Greek datasets to develop our $\lambda$-selection method. We start by exploring whether the model is misspecified for these data using PPC. We fit the model using MCMC, and obtain posterior samples $(\phi, \psi)^{\lambda, n}, n=1,\dots,N,$ given observed data $W^\text{obs}$. To generate replicated data $W^{\text{rep},\lambda,n}$, for each snippet $d \in \{1,\dots,D\}$, we first simulate the sense assignment $z_d^{\text{rep},\lambda,n} \sim p \left(\cdot | W_d^\text{obs}, \phi^{\lambda,n}, \psi^{\lambda,n} \right)$ with $p(\cdot | W_d, \phi, \psi)$ as in \eqref{eq:z_d_posterior}. Then, we simulate context words $w_{d,i}^{\text{rep},\lambda,n} | (z_d^{\text{rep},\lambda,n} = k) \sim \tilde{\psi}^{{\lambda,n} \atop {k,\tau_d}}$ independently for each position $i$ in $W_d^{\text{rep},\lambda,n}$. For the diagnostic function $s(W,\phi,\psi)$, we use the log likelihood $\log p(W|\phi,\psi)$, with $p(W|\phi,\psi)$ as in \eqref{eq:likelihood}, which gives the best results out of all the options explored. (The other options included the chi-squared statistic based on observed and expected counts, and various divergences between the parameters and the empirical distribution.)

\begin{figure}[t]
\centering
\makebox[\textwidth][c]{
\includegraphics[width = 1.1\textwidth, keepaspectratio]{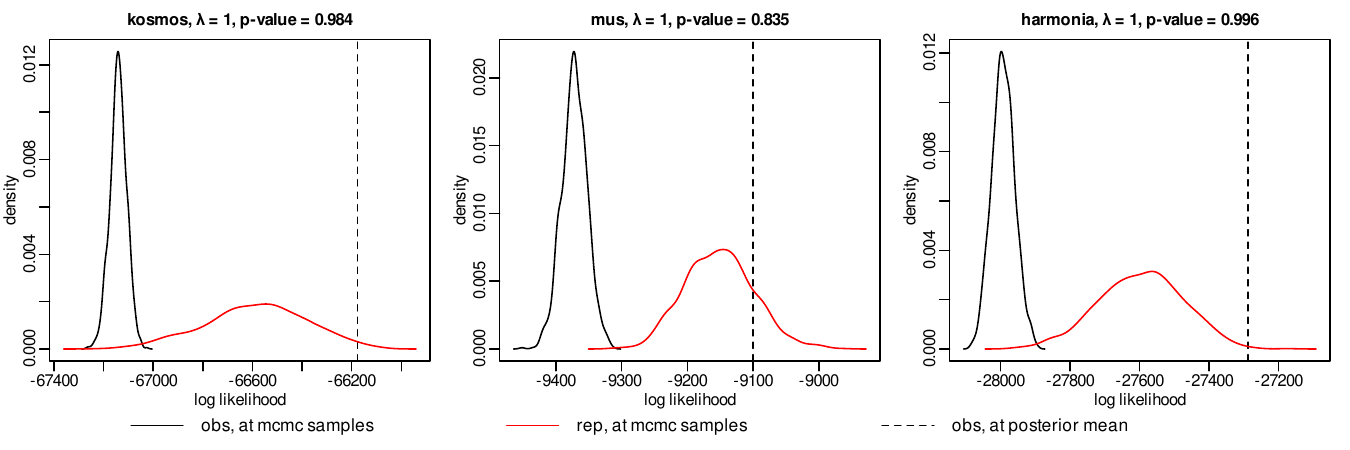}
}
\vspace{-20pt}
\caption{Posterior predictive checks on development data for $\lambda=1$, showing the distribution of the log-likelihood diagnostic}
\label{fig:PPC_dev_data_lambda_1}
\end{figure}

Figure~\ref{fig:PPC_dev_data_lambda_1} shows the PPCs in the $\lambda=1$ case. The solid black line shows the distribution of $\log p \left( W^{\text{obs}} | (\phi,\psi)^{\lambda} \right)$ computed on observed data at posterior samples; the red line shows the reference distribution of $\log p \left( W^{\text{rep},\lambda} | (\phi,\psi)^{\lambda} \right)$ computed on replicated data at posterior samples; and the dashed black line shows the point-value $\log p \left( W^{\text{obs}} | (\bar{\phi}, \bar{\psi})^{\lambda} \right)$ computed on observed data at the posterior means $(\bar{\phi}, \bar{\psi})^{\lambda}$. (Note that we actually take posterior means for the softmax-transformed probabilities $(\tilde{\phi}, \tilde{\psi})^{\lambda}$, but write $(\bar{\phi}, \bar{\psi})^{\lambda}$ to keep the notation tidy.) 
For all three datasets, the observed $\log p \left( W^{\text{obs}} | (\phi,\psi)^{\lambda} \right)$ lies well to the left relative to the reference distribution, suggesting that the individual posterior samples do not describe the observed data very well. On the other hand, $\log p \left( W^{\text{obs}} | (\bar{\phi}, \bar{\psi})^{\lambda} \right)$ lies to the right in all cases, suggesting that the posterior means tend to overfit the observed data, at least for ``kosmos'' and ``harmonia''; so there is potential misspecification, possibly due to the unmodelled latent correlations mentioned in Section~\ref{sec:model_and_inference}. 
As discussed at the end of Section~\ref{sec:PPCs}, we use the test statistic $\log p \left( W^{\text{obs}} | (\bar{\phi}, \bar{\psi})^{\lambda} \right)$ based on posterior means, and do not refer to $\log p \left( W^{\text{obs}} | (\phi,\psi)^{\lambda} \right)$ henceforth in this section.

\begin{figure}[!t]
\centering
\makebox[\textwidth][c]{
\includegraphics[width = 1.1\textwidth, keepaspectratio]{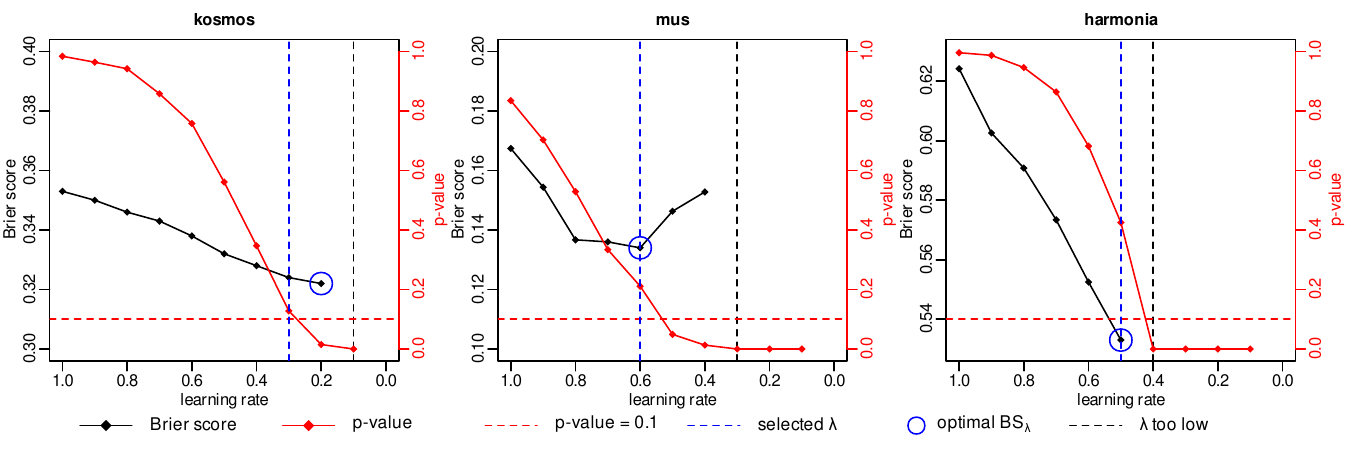}
}
\vspace{-20pt}
\caption{Brier scores and $p$-values for varying learning rates on development data}
\label{fig:BS_vs_pvalue}
\end{figure}

\begin{figure}[!t]
\centering
\makebox[\textwidth][c]{
\includegraphics[width = 1.1\textwidth, keepaspectratio]{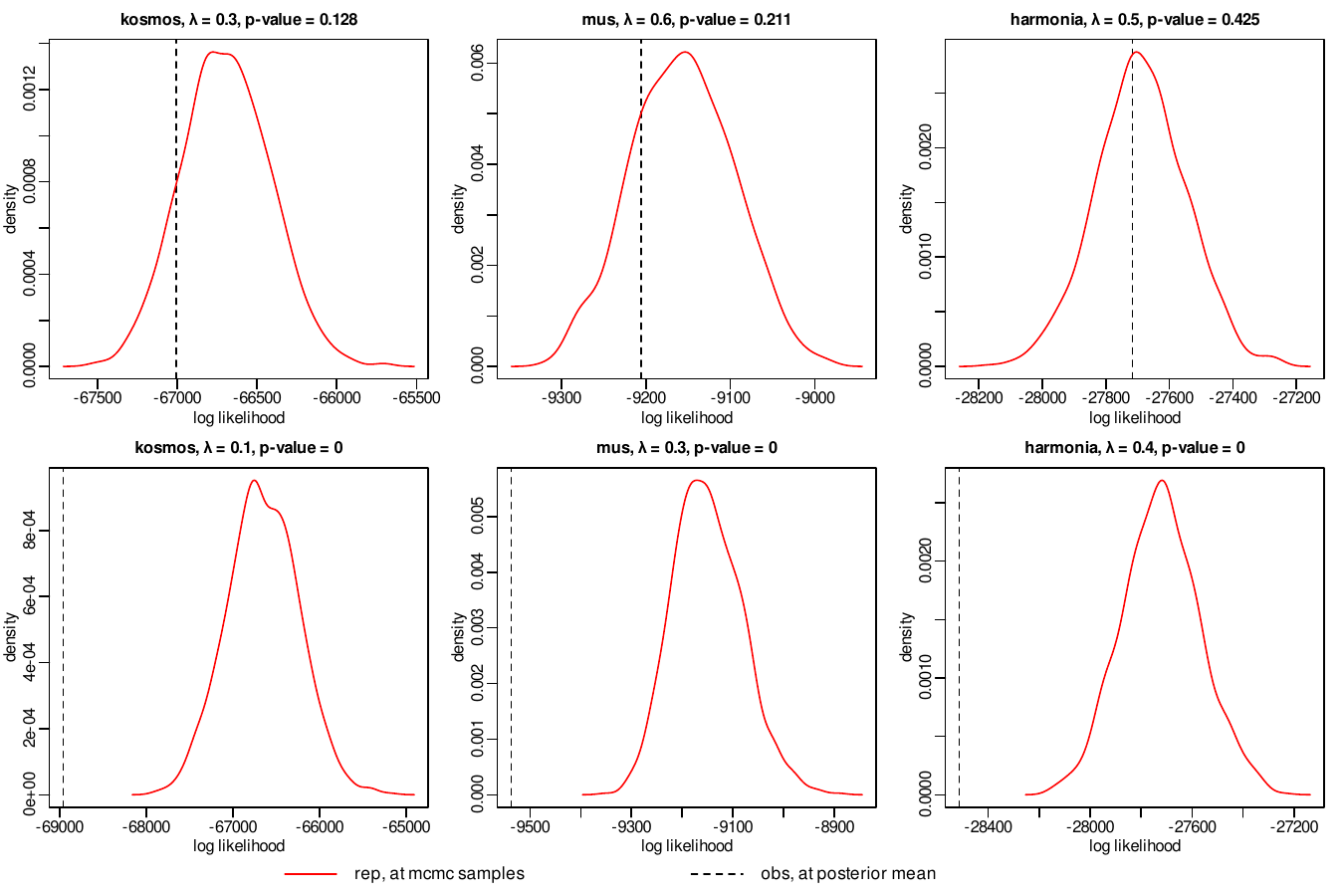}
}
\vspace{-20pt}
\caption{Posterior predictive checks on development data for $\lambda^\dagger = \argmin_{\lambda} \{p_\text{ppc}^\lambda: p_\text{ppc}^\lambda > 0.1\}$ (top row) and the $\lambda$ value at the point of collapse (bottom row)}
\label{fig:PPC_dev_data_lambda_2}
\end{figure}

We now examine the behaviour of $p$-values $p_\text{ppc}^\lambda$ and Brier scores $\text{BS}_\lambda$ as we reduce the learning rate $\lambda$. This is shown in Figure~\ref{fig:BS_vs_pvalue} for $\lambda=1,0.9,\dots,0.1$. Firstly, we note that $p_\text{ppc}^\lambda$ (solid red line) always decreases with $\lambda$ until it hits zero. This corresponds to the observed diagnostic $\log p \left( W^{\text{obs}} | (\bar{\phi}, \bar{\psi})^{\lambda} \right)$ always shifting to the left relative to the reference distribution of $\log p \left( W^{\text{rep},\lambda} | (\phi,\psi)^{\lambda} \right)$, as is shown explicitly in Figure~\ref{fig:PPC_dev_data_lambda_2} for two different $\lambda$ values for each dataset. This is as we would expect, since decreasing $\lambda$ corresponds to a smaller weight being given to the likelihood in the generalised Bayes' posterior \eqref{eq:generalised_bayes_EDiSC}.

Secondly, we note that $\text{BS}_\lambda$ (solid black line in Figure~\ref{fig:BS_vs_pvalue}) decreases (i.e. improves) with $\lambda$ up to the point that it hits the optimal $\text{BS}_{\lambda^*}$ (blue circle). If we keep reducing $\lambda$ beyond this point, either $\text{BS}_\lambda$ starts to increase again (as for ``mus''), or the inferred parameters stop being meaningful (which we will refer to as `collapse' for the reasons given in the \nameref{sec:appendix}). The latter is guaranteed to happen in any case for a sufficiently small $\lambda$ when the prior dominates the tempered likelihood in \eqref{eq:generalised_bayes_EDiSC}. Brier score cannot be computed after collapse, since there is no mapping between modelled senses and true senses, so the $\text{BS}_\lambda$ line ends just before this point.

Finally, we note that as soon as the parameters collapse (dashed black line in Figure~\ref{fig:BS_vs_pvalue}), $p_\text{ppc}^\lambda$ immediately drops to zero. PPCs at the point of collapse are shown in the bottom row of Figure~\ref{fig:PPC_dev_data_lambda_2}, clearly showing the observed diagnostic $\log p \left( W^{\text{obs}} | (\bar{\phi}, \bar{\psi})^{\lambda} \right)$ at these $\lambda$ values to be an outlier within the reference distribution of $\log p \left( W^{\text{rep},\lambda} | (\phi,\psi)^{\lambda} \right)$. The converse is not true: as $p_\text{ppc}^\lambda \to 0$, collapse may not yet have happened (as for ``kosmos'' and ``mus''), and $\text{BS}_\lambda$ may still be improving (as for ``kosmos''). However, it appears that a very low value of $p_\text{ppc}^\lambda$ does correlate, albeit loosely, with an under-optimised posterior in some sense. This suggests that a cutoff point (i.e. significance level) $\alpha$ for $p_\text{ppc}^\lambda$ could be a suitable criterion for $\lambda$ selection. As discussed in Section~\ref{sec:setting_lambda}, a conservative approach would be appropriate due to the miscalibrated $p$-values, so we set the cutoff point reasonably high at $\alpha = 0.1$ (dashed red line). 
Out of the candidate $\lambda$ values, we then select the learning rate $\lambda^\dagger$ as per \eqref{eq:selected_lambda} (dashed blue line). For our development data, this $\lambda^\dagger$ gives the optimal $\text{BS}_{\lambda^*}$ for ``mus'' and ``harmonia'', and very close to the optimal $\text{BS}_{\lambda^*}$ for ``kosmos''. By design, the misspecification at these $\lambda^\dagger$ values is not significant under PPC, as shown in the top row of Figure~\ref{fig:PPC_dev_data_lambda_2}.

\subsection{Results on test data} \label{sec:results_on_test_data}

\begin{table}[t]
\caption{Brier scores $\text{BS}_\lambda$ and $p$-values $p_\text{ppc}^\lambda$ for candidate learning rates $\lambda$ on test data. Optimal scores $\text{BS}_{\lambda^*}$ are in blue. Scores $\text{BS}_{\lambda^\dagger}$ for the learning rates $\lambda^\dagger$ selected using our method are boxed. Missing values for $\text{BS}_\lambda$ indicate collapse. All values are accurate to 3 s.f.} \label{tab:BS_pvalues_test_data}
\centering
\begin{adjustbox}{max width=0.9\textwidth}
\begin{tabular}{ S[table-format=1.1] S[table-format=1.3] S[table-format=1.3] S[table-format=1.4] S[table-format=1.3] S[table-format=1.3] S[table-format=1.3] S[table-format=1.4] S[table-format=1.3] S[table-format=1.3] S[table-format=1.3] }
\toprule
 & \multicolumn{2}{c}{bank split 1} & \multicolumn{2}{c}{bank split 2} & \multicolumn{2}{c}{bank split 3} & \multicolumn{2}{c}{bank split 4} & \multicolumn{2}{c}{bank split 5} \\
$\lambda$ & $\text{BS}_\lambda$ & $p_\text{ppc}^\lambda$ & $\text{BS}_\lambda$ & $p_\text{ppc}^\lambda$ & $\text{BS}_\lambda$ & $p_\text{ppc}^\lambda$ & $\text{BS}_\lambda$ & $p_\text{ppc}^\lambda$ & $\text{BS}_\lambda$ & $p_\text{ppc}^\lambda$ \\
\midrule
1   & 0.135 & 0.863 & 0.115 & 0.832 & 0.106 & 0.833 & 0.144 & 0.882 & 0.189 & 0.833 \\
0.6 & 0.128 & 0.428 & 0.111 & 0.393 & 0.103 & 0.412 & 0.139 & 0.459 & 0.178 & 0.374 \\
0.5 & 0.125 & 0.256 & 0.110 & 0.233 & 0.103 & 0.258 & 0.138 & 0.266 & 0.173 & 0.213 \\
0.4 & \boxed{0.124} & 0.114 & \boxed{0.109}\phantom{0} & 0.119 & \boxed{\textcolor{blue}{0.103}} & 0.126 & \boxed{0.138}\phantom{0} & 0.138 & \boxed{0.168} & 0.105 \\
0.3 & 0.121 & 0.045 & \textcolor{blue}{0.108\phantom{0}} & 0.030 & 0.104 & 0.036 & \textcolor{blue}{0.137\phantom{0}} & 0.048 & 0.162 & 0.030 \\
0.2 & \textcolor{blue}{0.118} & 0.009 & 0.109 & 0.009 & 0.106 & 0.007 & 0.137 & 0.009 & 0.158 & 0.007 \\
0.1 & 0.122 & 0     & 0.113 & 0     & 0.113 & 0.001 & 0.141 & 0     & \textcolor{blue}{0.152} & 0     \\
\toprule
 & \multicolumn{2}{c}{chair} & \multicolumn{2}{c}{apple} & \multicolumn{2}{c}{gay} & \multicolumn{2}{c}{mouse} & \multicolumn{2}{c}{bug} \\
$\lambda$ & $\text{BS}_\lambda$ & $p_\text{ppc}^\lambda$ & $\text{BS}_\lambda$ & $p_\text{ppc}^\lambda$ & $\text{BS}_\lambda$ & $p_\text{ppc}^\lambda$ & $\text{BS}_\lambda$ & $p_\text{ppc}^\lambda$ & $\text{BS}_\lambda$ & $p_\text{ppc}^\lambda$ \\
\midrule
1   & 0.140 & 0.656 & 0.0510 & 0.981 & 0.343 & 0.776 & 0.0474 & 0.968 & 0.300 & 0.974 \\
0.6 & \boxed{0.133} & 0.134 & 0.0496 & 0.899 & 0.299 & 0.217 & 0.0382 & 0.796 & 0.268 & 0.567 \\
0.5 & 0.129 & 0.037 & 0.0488 & 0.794 & \boxed{0.291} & 0.111 & 0.0359 & 0.692 & \boxed{0.267} & 0.222 \\
0.4 & 0.126 & 0.009 & 0.0484 & 0.652 & \textcolor{blue}{0.282} & 0.021 & 0.0348 & 0.474 & \textcolor{blue}{0.265} & 0.066 \\
0.3 & 0.124 & 0 & 0.0479 & 0.403 & 0.295 & 0 & \boxed{\textcolor{blue}{0.0335}} & 0.245 &  & 0 \\
0.2 & \textcolor{blue}{0.120} & 0 & \boxed{0.0467} & 0.208 &  & 0 & 0.0337 & 0.077 &  & 0 \\
0.1 & 0.125 & 0 & \textcolor{blue}{0.0461} & 0.029 &  & 0 & 0.0393 & 0.004 &  & 0 \\
\bottomrule
\end{tabular}
\end{adjustbox}
\end{table}

We now give the results of applying our $\lambda$-selection method on the 10 test datasets in Table~\ref{tab:BS_pvalues_test_data}. For the five split ``bank'' datasets, our method consistently selects $\lambda^\dagger = 0.4$. The $\text{BS}_{\lambda^\dagger}$ for the selected rate is optimal, or almost optimal, for splits 2--4. For splits 1 and 5, $\text{BS}_{\lambda^\dagger}$ is still closer to the optimal $\text{BS}_{\lambda^*}$ than $\text{BS}_1$, which is a successful outcome as per our criteria set out at the end of Section~\ref{sec:assessing_performance}. Furthermore, the fact that our method selects the same $\lambda^\dagger$ for all five splits makes sense: the split datasets are a random partition of a larger dataset, and we expect misspecification to be similar in nature and degree across partitions. The learning-rate estimates are robust to other details of the data. 

Out of the other five test datasets, for all except the ``chair'' data, our method again selects a rate $\lambda^\dagger$ that is either optimal or almost optimal as measured by Brier score. Interestingly, $p_\text{ppc}^\lambda$ for the ``apple'' and ``mouse'' data drops more slowly with decreasing $\lambda$ compared to the other datasets, and our method selects relatively lower $\lambda^\dagger$ values for these data. As mentioned in Section~\ref{sec:dev_and_test_data}, these data samples are subsets of snippets selected from the corpus to ensure adequate representation of both target-word senses. The selection mechanism tends to select snippets with strong and clearly evidenced meanings, and as a consequence the model senses in the posterior are more sharply separated. Hence, $\lambda$ needs to be reduced more to have enough of a `softening' effect on the sense separation. 

On the ``chair'' data, the $\text{BS}_{\lambda^\dagger}$ returned is an improvement on $\text{BS}_1$, though it is not as close to the optimal $\text{BS}_{\lambda^*}$ as in the other examples. Recall (from Section~\ref{sec:assessing_performance}) that we are using predictive accuracy as a criterion for performance measurement only for the sake of objectivity, since we do not know the true DGP. However, the goals of predictive accuracy and true-model recovery are not always in sync (witness the contrasting consistency behaviour of AIC and BIC). GBI attempts to correct model misspecification, and is therefore more aligned with the latter than the former. It seems that the two goals are less synchronised for the ``chair'' data than the other datasets. In any case, as set out in the introduction, our method is not guaranteed to find the optimal model --- we promote it as an intuitively well-founded and computationally efficient way to select $\lambda$, and one that works well in all of our test cases.

\section{Discussion} \label{sec:discussion}

In this exploratory work, we considered the problem of selecting an appropriate learning rate $\lambda$ for GBI, within the specific context of the recent EDiSC model that is used to quantify changes in target-word senses over time. We argued that it makes intuitive sense to set the learning rate using PPC based on the log likelihood diagnostic, such that we select the lowest rate where a PPC is not rejected at the (nominal) 10\% significance level. This approach is computationally efficient and can be readily implemented with MCMC sampling. We developed this approach using experiments on three datasets, and tested it on 10 new datasets. We found the approach to work very well in all cases as measured by predictive accuracy quantified using Brier scores: the accuracy attained with a learning rate selected using our method is very close to, or exactly at, the optimal level. Some further insights into the mechanism by which tempering helps are given in the \nameref{sec:appendix}.

Our $\lambda$-selection method is adapted to our model and training data; and although it is effective in our setting, it is not universally applicable. The method seems well suited to cases where the posterior mean overfits the data, perhaps due to unmodelled latent correlations in the data. In these cases, for $\lambda=1$ (i.e. standard Bayes), the agreement score $s(y^{\text{obs}}, \bar{\vartheta}^\lambda)$ for observed data at the posterior mean is typically higher than the agreement scores $s(y^{\text{rep},\lambda,n}, \vartheta^{\lambda, n}), n=1,\dots,N,$ for replicated data at posterior samples. As $\lambda$ decreases, we remove information entering the analysis from the data, and the fitted agreement score $s(y^{\text{obs}}, \bar{\vartheta}^\lambda)$ reduces towards the replicated agreement scores. This reflects the fact that the model is over-valuing the information in the data: the data contains latent unmodelled correlations, so its effective sample size is lower than its nominal sample size. 

We ran additional tests of the method for simple Bayesian linear models, using simulated small datasets and/or low-dimensional parameters, and found that the method was not useful for selecting $\lambda$. In these tests, the property mentioned above was not present by construction. Therefore, the fitted agreement score $s(y^{\text{obs}}, \bar{\vartheta}^\lambda)$ at $\lambda=1$ was in a random location relative to the replicated agreement scores $s(y^{\text{rep},\lambda,n}, \vartheta^{\lambda, n}), n=1,\dots,N$. 
Taking $\lambda<1$ only decreases the fitted agreement, potentially moving it further away from the replicated agreement scores. In future work, we would like to experiment further, using simulated data designed more accurately to mimic the misspecification seen in our setting, in order to provide other use cases. 

The reasons for the behaviour described above are not obvious. However, it is clear that a reasonably large dataset is required to see this behaviour, and a high-dimensional parameter space seems warranted (which is a natural consequence of the data in our setting). The posterior should display a complex underlying correlation structure, so that individual posterior samples $\vartheta^{\lambda,n},n=1,\dots,N,$ do not describe the data well but the posterior \textit{mean} $\bar{\vartheta}^\lambda$ does so. This corresponds, at $\lambda=1$, to the agreement score $s(y^{\text{obs}}, \bar{\vartheta}^\lambda)$ at the posterior mean being higher than the equivalent agreement scores $s(y^{\text{obs}}, \vartheta^{\lambda,n})$ at the posterior samples. As seen in Figure~\ref{fig:PPC_dev_data_lambda_1}, this is the case in our real-data setting. However, in the additional tests mentioned above where our $\lambda$-selection method was not useful, $s(y^{\text{obs}}, \bar{\vartheta}^\lambda)$ was generally representative of $s(y^{\text{obs}}, \vartheta^{\lambda,n})$ at the posterior samples.

Secondly, the likelihood equation \eqref{eq:likelihood} plays a role. The actual data, as discussed in Section~\ref{sec:model_and_inference}, exists in pairs with a snippet $W_d$ and a sense $z_d$ for all $d=1,\dots,D$. However, the latent sense assignments $z$ are not observed, so \eqref{eq:likelihood} is computed marginally over $z$. 
In our experience, GBI methods have most to offer when the model has a high-dimensional latent parameter that is misspecified in the model, which is certainly the case here. Each snippet brings with it one of these missing $z_d$, so they are numerous. The model plays an important role in weighting sense assignments, as any $W_d$ itself may contain relatively little information about $z_d$. Hence, if the model is misspecified, methods treating misspecification have a role as well. On the other hand, if the data were strongly informative of the sense, this would not be the case.

Thirdly, the power $\lambda$ in \eqref{eq:generalised_bayes_EDiSC} goes outside the sum over possible sense assignments in \eqref{eq:likelihood}, so the power is on the partially marginalised likelihood. In this sum, if the fit is good, the term for the correct sense assignment would be largest. Taking the power \textit{outside} the sum has the effect of `protecting' the largest term in the sum when reducing $\lambda$.

Whilst the above explanations give some intuition as to when our $\lambda$-selection approach might work, we have not yet defined the exact conditions required. However, the systematic nature of results seen in all the real-data examples studied in this paper, and the success of our method in all cases, suggests that this line of investigation is worth exploring further.

\section*{Implementation}
The code and data used to produce the results reported in this paper are available from \url{https://github.com/schyanzafar/GBI}.

\appendix
\section*{Appendix} \label{sec:appendix}
\section*{Analysis of generalised Bayes' posterior} \label{sec:analysis}

To gain some insight into the mechanism by which tempering the likelihood improves model performance in our setting, it helps to analyse what happens to the inferred parameters $(\tilde{\phi}, \tilde{\psi})^\lambda$ as we reduce $\lambda$. We show these analyses for ``mus'' only, but the results discussed here are also typical for ``kosmos'' and ``harmonia''.

\begin{table}[!t]
\caption{Top 10 context words under each model sense of ``mus'' for different learning rates. Words repeated across multiple senses are indicated in \textcolor{red}{red}.}
\label{tab:mus_top_words}
\centering
\begin{adjustbox}{max width=\textwidth}
\begin{tabular}{c l l l l l l l l l l}
\toprule
Sense & \multicolumn{10}{l}{Top 10 context words $\lambda=1$} \\
\midrule
1 & \textgreek{λέγω} & \textgreek{φημί} & \textgreek{γῆ} & \textgreek{γίγνομαι} & \textgreek{πᾶς} & \textgreek{εἶτα} & \textgreek{νῦν} & \textgreek{μῦς} & \textgreek{γαλέη} & \textgreek{πλῆθος} \\
2 & \textgreek{\textcolor{red}{νεῦρον}} & \textgreek{ὀστέον} & \textgreek{φλέψ} & \textgreek{βραχίων} & \textgreek{ἔχω} & \textgreek{τένων} & \textgreek{ἄρθρον} & \textgreek{μῦς} & \textgreek{πυρετός} & \textgreek{μυελός} \\
3 & \textgreek{σωλήν} & \textgreek{κτείς} & \textgreek{ὄστρεον} & \textgreek{χρύσεος} & \textgreek{λεπάς} & \textgreek{χήμη} & \textgreek{κόγχη} & \textgreek{ὄστρειον} & \textgreek{πίννα} & \textgreek{\textcolor{red}{νεῦρον}} \\
\toprule
Sense & \multicolumn{10}{l}{Top 10 context words $\lambda=0.6$} \\
\midrule
1 & \textgreek{λέγω} & \textgreek{φημί} & \textgreek{γῆ} & \textgreek{γίγνομαι} & \textgreek{πᾶς} & \textgreek{πολύς} & \textgreek{\textcolor{red}{ἔχω}} & \textgreek{πλῆθος} & \textgreek{εἶτα} & \textgreek{νῦν} \\
2 & \textgreek{νεῦρον} & \textgreek{ὀστέον} & \textgreek{φλέψ} & \textgreek{βραχίων} & \textgreek{\textcolor{red}{ἔχω}} & \textgreek{τένων} & \textgreek{ἄρθρον} & \textgreek{μῦς} & \textgreek{πυρετός} & \textgreek{κίνησις} \\
3 & \textgreek{σωλήν} & \textgreek{κτείς} & \textgreek{λεπάς} & \textgreek{χήμη} & \textgreek{ὄστρεον} & \textgreek{πίννα} & \textgreek{χρύσεος} & \textgreek{ὄστρειον} & \textgreek{κόγχη} & \textgreek{μῦς} \\
\toprule
Sense & \multicolumn{10}{l}{Top 10 context words $\lambda=0.3$} \\
\midrule
1 & \textgreek{\textcolor{red}{νεῦρον}} & \textgreek{\textcolor{red}{ὀστέον}} & \textgreek{\textcolor{red}{φλέψ}} & \textgreek{\textcolor{red}{λέγω}} & \textgreek{\textcolor{red}{φημί}} & \textgreek{μῦς} & \textgreek{\textcolor{red}{ἔχω}} & \textgreek{γίγνομαι} & \textgreek{γῆ} & \textgreek{πολύς} \\
2 & \textgreek{\textcolor{red}{νεῦρον}} & \textgreek{\textcolor{red}{ὀστέον}} & \textgreek{\textcolor{red}{φλέψ}} & \textgreek{μῦς} & \textgreek{\textcolor{red}{βραχίων}} & \textgreek{\textcolor{red}{ἔχω}} & \textgreek{\textcolor{red}{λέγω}} & \textgreek{\textcolor{red}{φημί}} & \textgreek{ἄρθρον} & \textgreek{ὦμος} \\
3 & \textgreek{\textcolor{red}{νεῦρον}} & \textgreek{μῦς} & \textgreek{σωλήν} & \textgreek{κτείς} & \textgreek{\textcolor{red}{φημί}} & \textgreek{\textcolor{red}{λέγω}} & \textgreek{λεπάς} & \textgreek{\textcolor{red}{ὀστέον}} & \textgreek{\textcolor{red}{βραχίων}} & \textgreek{πίννα} \\
\bottomrule
\end{tabular}
\end{adjustbox}
\end{table}

As discussed in \citet[Section~5.2]{EDiSC_ZAFAR2024108011}, a natural way to examine the model output is to look at the context words with the highest probabilities under each model sense, marginally over time, using the posterior mean $\bar{\psi}^\lambda$. We show the top 10 context words under the model fits for $\lambda=1$ (standard Bayes), $\lambda=0.6$ (optimal) and $\lambda=0.3$ (collapsed) in Table~\ref{tab:mus_top_words}. With some knowledge of ancient Greek, or with the help of a dictionary (e.g. Wiktionary), or by comparing against expert annotation, the three model senses in the first two cases can be identified as mouse, muscle and mussel respectively. The output for $\lambda=1$ and $\lambda=0.6$ is quite similar, and the model senses retain their separate identities in both cases. However, for $\lambda=0.3$, the model senses are no longer distinguishable from each other, as they are displaying very similar sets of top words; or, in other words, they have `collapsed'. They do, nevertheless, still reflect the overall context-word probabilities (across all senses) learnt from the data. If we keep reducing $\lambda$, eventually all model senses look the same, and all context words take uniform probabilities as per the prior. 

\begin{figure}[!t]
\centering
\includegraphics[width = 1\textwidth, keepaspectratio]{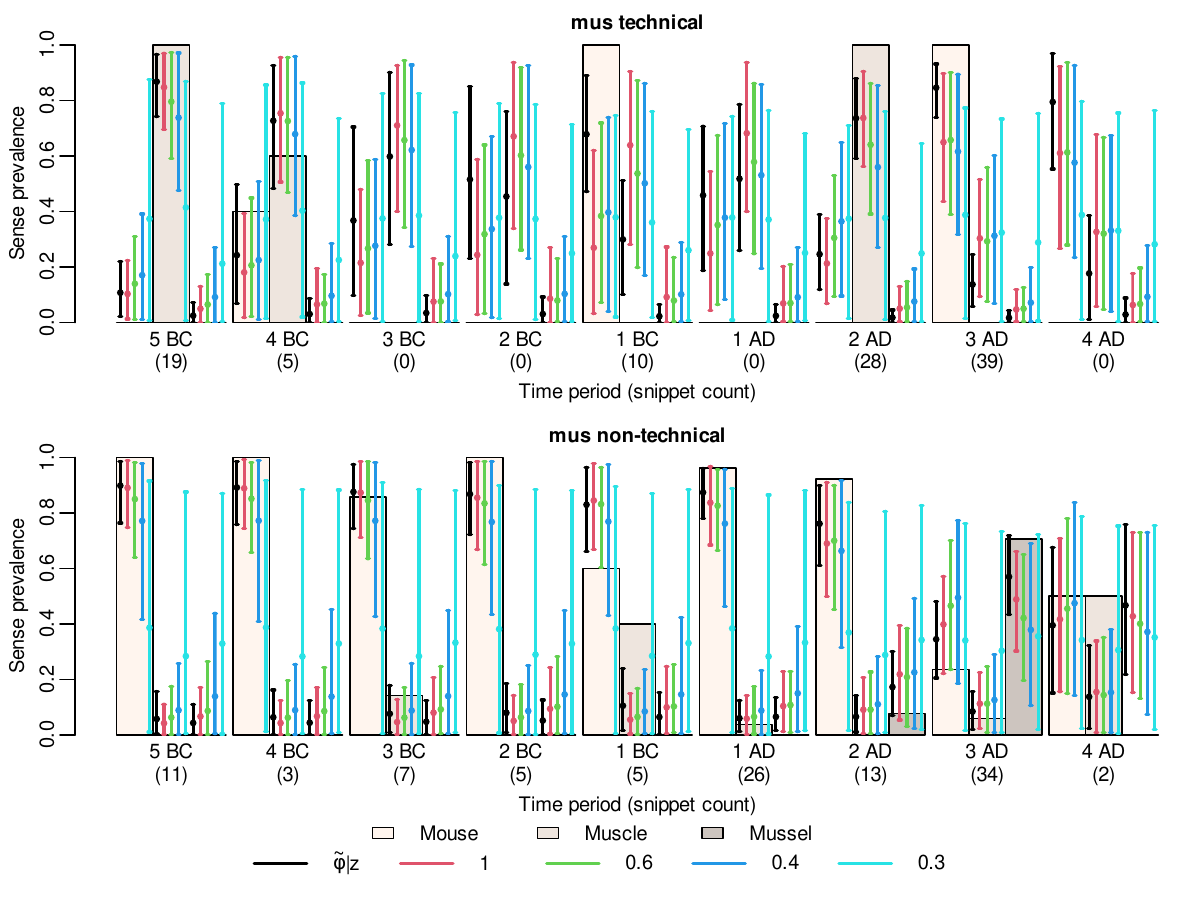}
\vspace{-20pt}
\caption{95\% HPD intervals (error bars) and posterior means (circles) for the model output on ``mus'' data. Results are shown for the proxy truth $\tilde{\phi}|z$ and the posteriors $\tilde{\phi}^\lambda|W$ for $\lambda \in \{1, 0.6, 0.4, 0.3\}$. Expert-annotated empirical sense prevalence (solid bars) for the three true senses is also shown.}
\label{fig:mus_phi_error_bars}
\end{figure}

Sense prevalence $\tilde{\phi}$ is a lower-dimensional object reflecting the behaviour of higher-dimensional senses $\tilde{\psi}$ (where each sense $k$ for time $t$ is a distribution $\tilde{\psi}^{k,t}$ over context words $1,\dots,V$) and is therefore easier to visualise. We argued in Section~\ref{sec:assessing_performance} that a comparison of model posterior $\tilde{\phi}^\lambda|W$ against the proxy `true' model posterior $\tilde{\phi}|(z=o)$ should not be used to measure performance. However, this comparison does help us analyse the posterior behaviour, and is shown in Figure~\ref{fig:mus_phi_error_bars} for several learning rates. We see that, as we reduce $\lambda$ from 1 to 0.6, the effect is a slight increase in variance without much change in location. This is the same effect that makes likelihood tempering useful in an MCMC convergence context (cf. \citealt[Appendix~C]{EDiSC_ZAFAR2024108011}). In this case, if $\tilde{\phi}^1|W$ is slightly misspecified in the sense of not achieving enough overlap with the proxy-true $\tilde{\phi}|(z=o)$, the increased variance allows $\tilde{\phi}^{0.6}|W$ to correct the misspecification by covering more of the posterior space and increasing the overlap. As a consequence, this correction improves overall predictive accuracy on average. 

When we reduce $\lambda$ further to 0.4, the posterior location starts to gets less accurate due to lower influence from the data, and the variance increases to cover regions of the posterior space that do not overlap with the proxy truth. Thus, whilst parts of the posterior still return good predictive accuracy, the overall accuracy reduces on average. However, up to this point, the model may still be described as somewhat `well specified' in the sense of achieving reasonable overlap with the proxy-true $\tilde{\phi}|(z=o)$. Reducing $\lambda$ further to 0.3, we see that the posterior $\tilde{\phi}^\lambda|W$ now becomes too diffuse, and too uniform across senses, to be described as well specified in any way. This explains why a reduction in $\lambda$ helps up to a point, and why the degree of misspecification (which we measure using PPC) is an intuitive choice to demarcate this point.

\bibliography{bibliography.bib}

\end{document}